**Citation algorithms for identifying research milestones driving biomedical innovation**

Jordan A. Comins[a,*] and Loet Leydesdorff[b]

___________________________

[a] *corresponding author*; Center for Applied Information Science, Virginia Tech Applied Research Corporation, Arlington, VA, United States; jcomins@gmail.com
[b] Amsterdam School of Communication Research (ASCoR), University of Amsterdam, PO Box 15793, 1001 NG Amsterdam, The Netherlands; loet@leydesdorff.net

**Abstract**

Scientific activity plays a major role in innovation for biomedicine and healthcare. For instance, fundamental research on disease pathologies and mechanisms can generate potential targets for drug therapy. This co-evolution is punctuated by papers which provide new perspectives and open new domains. Despite the relationship between scientific discovery and biomedical advancement, identifying these research milestones that truly impact biomedical innovation can be difficult and is largely based solely on the opinions of subject matter experts. Here, we consider whether a new class of citation algorithms that identify seminal scientific works in a field, Reference Publication Year Spectroscopy (RPYS) and multi-RPYS, can identify the connections between innovation (e.g., therapeutic treatments) and the foundational research underlying them. Specifically, we assess whether the results of these analytic techniques converge with expert opinions on research milestones driving biomedical innovation in the treatment of Basal Cell Carcinoma. Our results show that these algorithms successfully identify the majority of milestone papers detailed by experts (Wong and Dlugosz 2014) – thereby validating the power of these algorithms to converge on independent opinions of seminal scientific works derived by subject matter experts. These advances offer an opportunity to identify scientific activities enabling innovation in biomedicine.

**Introduction**

Biomedical innovation is guided, at least in part, by historical influences (Agarwal and Searls, 2009). In some cases, biomedical innovation is guided by key scientific discoveries, referred to here as research milestones treatments (Wong and Dlugosz, 2014). Nelson, Buterbaugh, Perl, & Gelijns (2011) distinguish analytically among three enabling forces in medical innovation: advances of scientific understanding of diseases, learning in clinical practices, and advances in technological capabilities (which often originate outside of medicine) for the development of novel modalities of diagnosis and treatment. However, interactions between supply-side factors and demand originating from diseases can be expected to shape co-evolutions along trajectories enabled by technological capabilities (Petersen, Rotolo, & Leydesdorff, 2016). From the historiographic perspective in innovation studies, however, the innovation itself is focal (Von Hippel, 1988); but the scientific knowledge base of innovations can be excavated from the literature (Garfield, Sher, & Torpie, 1964). Medical innovations are sometimes triggered by breakthroughs and new developments on the supply side (Leydesdorff & Rafols, 2011; Mina, Ramlogan, Tampubolon, & Metcalfe, 2007).

Connecting major advances in medical treatments with earlier research milestones is considered essential for public appreciation of the role of basic research discoveries in major health advances (Williams et al., 2015) and offers an opportunity to more fully consider the factors leading to new drugs (Nelson, Buterbaugh, Perl, & Gelijns, 2011; Swanson, 1990). Despite the value of the translation of scientific activities into medical innovations (Mogoutov, Cambrosio, Keating, & Mustar, 2008), current approaches for retrieving research milestones require considerable time and commitment of subject matter experts. In this brief communication, we assess the utility of using computational approaches to identify research milestones, which may allow for broader engagement of scientific and non-scientific communities with a field's intellectual history.

Eugene Garfield, who first created the *Science Citation Index,* found the development of citation algorithms to reconstruct the intellectual history of scientific fields a top priority for scientometrics (e.g., Garfield et al., 1964). Citations serve as a functional linkage between ongoing scientific efforts with prior endeavors ( Price, 1965; Garfield, Malin, & Small, 1978; Radicchi, Fortuno, & Castellano, 2008). The accumulation of citations is therefore commonly used as a proxy for quantifying the scholarly impact of research (Kostoff and Shlesinger 2005; Marx et al. 2014; van Raan 2000).

Recent years have witnessed noteworthy advances in citation algorithms for identifying the seminal works within the history of a research field. In particular, the development of a new technique, known as Reference Publication Year Spectroscopy (RPYS), offers the possibility of visualizing key seminal papers for a research field using publicly available analysis tools (Comins and Hussey 2015a; Comins and Leydesdorff 2016a; Leydesdorff et al. 2014; Marx et al. 2014; Thor et al. 2016).[1] Up until now, use of RPYS for identifying seminal works has been limited to scientific fields where the authors can consider themselves as experts (Comins and Hussey 2015a, 2015b; Comins and Leydesdorff 2016a; Elango et al. 2016; Marx and Bornmann 2013; Marx et al. 2014; Wray and Bornmann 2015) and can therefore support the interpretation of results. Ultimately, this means that testing and evaluation of RPYS-based algorithms suffers from a lack of objectivity as the authors themselves also served as "referees" of the quality of RPYS results. Despite the potential of this technique for identifying seminal works, no studies on RPYS to date have *compared* its performance to the historical account of research milestones developed *independently* by experts.

---

[1] Implementations of RPYS can be retrieved at http://www.crexplorer.net/ and http://comins.leydesdorff.net/ . CRExplorer (the first option) offers the additional option of disambiguation of the cited references (Thor et al., 2016), while RPYS i/o (the second option) offers in addition to standard RPYS, Mult-RPYS as a graphically rich visualization (Comins & Leydesdorff, 2016). A pilot version of this same technique (rpys.exe) can be found at http://leydesdorff.net/software/rpys/ (Marx, Bornmann, Barth, & Leydesdorff, 2014).

This study represents a first attempt to evaluate RPYS performance more objectively by comparing the results of the technique to seminal articles already identified within a research area by subject matter experts. Here, we utilize a recent article published in the *Journal of Investigative Dermatology* to compare the milestones identified by subject matter experts on basal cell carcinoma (Wong and Dlugosz 2014) to those identified via RPYS analyses. Demonstrating that RPYS analyses converge on a similar set of milestones as those identified by experts offers the possibility that such tools can support a broader range of needs in science, including rapid identification of truly landmark findings in a field and informing students and non-academics about pivotal discoveries in the history of research areas.

**Methods**

To assess the performance of RPYS in identifying key milestones described by Wong and Dlugosz (2014), we began by devising search queries capturing the main topics discussed by their article. Search strings were generated by using terms found in the title of the paper as well as within the main text. Our goal was to make a series of searches that progressively become more complex – this would enable us to assess the effectiveness of the algorithms in identifying milestones for straightforward searches as well as more sophisticated search strings. Our search strings are shown in Table 1.

| Web of Science Search | Records | References |
|---|---|---|
| A. TS=(basal cell carcinoma* hedgehog) | 1148 | 23,644 |
| B. TS=(hedgehog inhibit* treatment* basal cell carcinoma*) | 244 | 6,920 |
| C. TS=((cyclopia hedgehog) OR (cyclopamine hedgehog)) | 918 | 20,131 |
| D. TS=(basal cell carcinoma* smoothened antagonist) | 92 | 2,700 |

**Table 1.** Web of Science search queries used to identify records related to basal cell carcinoma, the hedgehog signaling pathway and potential treatments.

For instance, search A was a topic search for "basal cell carcinoma* hedgehog", and was the broadest in terms of number of records retrieved and the number of references they cite. These terms were taken directly from the first four words of the title of Wong and Dlugosz (2014), "**Basal Cell Carcinoma, Hedgehog** Signaling, and Targeted Therapeutics: The Long and Winding Road." Search B focused the results from A by including terms related to treatment pathway inhibition by performing a topic search for "hedgehog inhibit* treatment* basal cell carcinoma*". These terms were found in both the title and subtitles of the paper. Search C was a topic search for "(cyclopia and hedgehog) or (cyclopamine and hedgehog)", thereby covering the relationship between the hedgehog signaling pathway and its relationship to cyclopamine, a teratogen that can cause the birth defect cyclopia. These search terms were extracted from the title of the article, some of the subtitles and some of the main body text. The terms cyclopia and cyclopamine occurred frequently throughout the document (including figures and references), 18 times. Search D was the narrowest, performing a topic search of "basal cell carcinoma* smoothened antagonist". This search describes the treatment of basal cell carcinoma by the drug Vismodegib, a competitive antagonist of the smoothened receptor, which was mentioned in the article's main body text and in its figures. For our initial assessment of RPYS methods to identify milestones, the data analyzed represented the union of Searches A-D; for our second assessment, Searches A-D were analyzed individually. Data was downloaded from the Thomson Reuters Web of Science (WoS) in February 2016.

Earlier work provides details on the procedures for performing standard RPYS (Comins and Hussey 2015b; Elango et al. 2016; Leydesdorff et al. 2014; Marx and Bornmann 2013; Marx et al. 2014; Thor et al. 2016; Wray and Bornmann 2015). In brief, the standard RPYS procedure aggregates cited references from a set of citing publications. For each year, the sum of the number of cited references are plotted by their publication year. This data is then de-trended by taking the difference between the number of cited references for any given reference publication year (e.g., $n$) from the 5-year median of reference publications ($y$-2, $y$-1, $y$, $y+1$, $y+2$). Visualizations were produced using procedures similar to those described in Comins &

Leydesdorff (2016b). Peak years appear as those where the de-trended value of cited references in a given year is higher than those of neighboring years. Multi-RPYS extends the impact of these techniques by first segmenting the data in terms of the publication years of the citing sets, performing the standard RPYS analysis within each set and then rank transforming the de-trended results to compare influential references across the history of the citing set (Comins & Leydesdorff, 2016a; Comins & Leydesdorff, 2016b).

To determine the performance of RPYS in identifying Wong and Dlugosz's milestones (2014; figure 2), we defined success by two criteria. First, our RPYS results needed to reveal a *peak year* corresponding to the years shown in Wong and Dlugosz milestones figure. Second, these *peak years* must list the key articles described by Wong and Dlugosz as one of the 10 most referenced articles in that year. Use of CRExplorer (available at http://www.crexplorer.net) to clean references enhanced our ability to determine whether a seminal milestone paper was indeed amongst the 10 most referenced articles in a given year (Thor et al., 2016). Without disambiguation of cited reference, attributing reference counts to articles could yield noisy results – leveraging CRExplorer to disambiguate cited references helps to mitigate this concern. Finally, visualizations were generated using the online tool RPYS i/o (at http://comins.leydesdorff.net), which conducts both standard and multiple RPYS analyses on WOS datasets.

**Multi-RPYS for identifying years with research milestones**

Using four informed searches (Table 1), we retrieved between 92 and 1,148 records at the Web of Science and these records contained between 2,700 to 23,644 cited references. Across the 2,402 total records retrieved from these four searches, 1,948 were unique records with 33,481 references (after *disambiguation* of cited references using clustering methods; see Thor et al, 2016). Multi-RPYS analysis of these 1,948 records (shown in Figure 1) visually reveal multiple bands when potentially influential references were published.

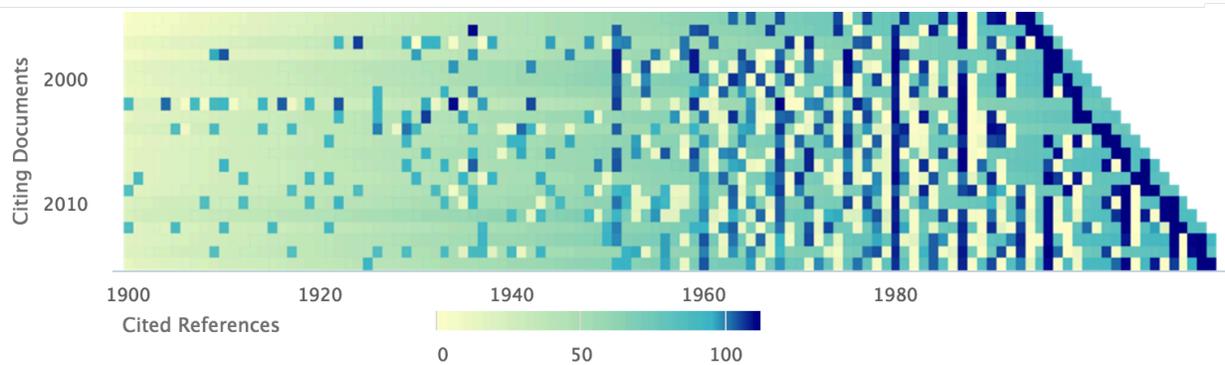

**Figure 1.** Multi-RPYS results for the union of the four searches outlined in Table 1. Cited References are placed along the x-axis and citing documents along the y-axis. Color represents the cell's ranked RPYS results (see methods) and bands, or columns, in the heatmap are demonstrative of persistently influential works. For example, the band in 1980, is driven by the Nobel prize winning work of Nusslein-Volhard and Wieschaus.

To assess which ones appear most *important*, we submitted results from multi-RPYS to inferential statistical tests to ascertain the cited years that appear most indicative of a milestone year. Confirming the intuition provided by our visualization in Figure 1, we found a significant difference across the cited reference years ($F = 12.94$; df = 113; $p < 0.0001$). From here, we assessed which cited reference years appear to deviate the most, indicative of a milestone year. We did so by utilizing a Tukey's HSD post-hoc test and identifying the ten cited reference years with the largest Least Square Means Differences: (1) 1980, (2) 2009, (3) 2012, (4) 1987, (5) 2004, (6) 1996, (7) 2011, (8) 2013, (9) 1968 and (10) 2006.

From this list, 70% matched the milestone years discussed by Wong and Dlugosz (2014). Interestingly, Wong and Dlugosz (2014) designate in total 25 milestones across 18 separate years from 1827 until 2012. Statistically, this would suggest that the chance of correctly identifying a milestone year from 1827-2012 by chance is approximately 10% (18 milestone years / 185 years under study). Our results using multi-RPYS far exceed chance expectations for identifying milestone years and suggests that the combination of multi-RPYS and inferential statistics can rapidly identify milestone years in the history of a research field. In the next section, we more thoroughly examine whether RPYS technique not only identify milestone years, but the precise articles noted by experts as seminal. To do so, we take a methical approach, using standard RPYS analyses on each of the four individual WoS record searches and describing their associated results.

## Standard RPYS for identifying specific milestone discoveries

In this section, each set of records from Searches A-D (Table 1) was submitted to its own RPYS analysis (Figure 2) and compared to the seminal papers discussed by Wong and Dlugosz (2014). In this context, seminal papers were determined by mapping publications from the authors' bibliography to the milestones described in their article (see Wong and Dlugosz, 2014; figure 2). In addition, several milestones were attributed to multiple publications, bringing a grand total of 30 seminal works described by Wong and Dlugosz (2014). Across the RPYS analyses of our four Web of Science searches we successfully capture 23 out of the 30, or 76%, of these milestone articles.

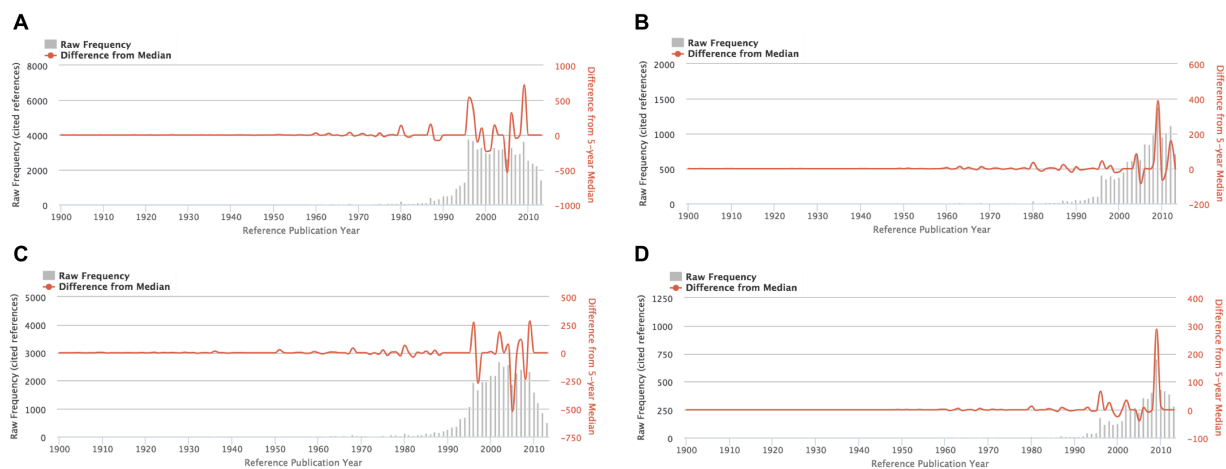

**Figure 2.** RPYS results for the four searches outline in Table 1. Panels A-D correspond to searches A-D described in Table 1, respectively.

Next, each peak coinciding with the date of a milestone described by Wong and Dlugosz (2014) was further inspected to determine the most referenced articles from that year. As noted in our methods section, RPYS analysis was determined to successfully identify a milestone article if (1) the spectrogram revealed a peak on the corresponding milestone year and (2) the milestone articles noted by Wong and Dlugosz were amongst the 10 most referenced articles from that year. These results are detailed in Table 2. Nearly every peak year corresponding to a milestone year contained a milestone article as the top referenced article.

| Milestones (Wong & Dlugosz, 2014; Figure 2) | Milestone Article (Wong & Dlugosz, 2014; bibliography) | RPYS Peak | Rank in Search | | | |
|---|---|---|---|---|---|---|
| | | | A | B | C | D |
| 1827: First clinical description of BCC | Jacob (1827). *Dublin Hosp Rep* | Yes | 1 | 1 | | |
| 1950s: One eyed lambs in Idaho | Binns et al (1959) *J Am Vet Med Res*[a] | Yes | | | 1 | |
| 1960: Nevoid BCC Syndrome described | Gorlin & Goltz (1960). *N Engl J Med* | Yes | 1 | 1 | | |
| 1963: Corn lily is teratogenic | Binns et al (1963). *Am J Vet Res* | Yes | 1 | 1 | 1 | 1 |
| 1968: Teratogen in corn lily identified: cyclopamine | No Document Referenced[b] | Yes | 1 | 1 | 1 | 1 |
| 1980: Hh & other embryonic patterning genes discovered | Nusslein-Volhard, Wieschaus (1980). *Nat* | Yes | 1 | 1 | 1 | 1 |
| 1992: *Drosophila* hh gene cloned | No Document Referenced | No | | | | |
| 1993: Mammalian Hh genes cloned | No Document Referenced[c] | Yes | | | | 1 |
| 1996: *PTCH1* mutations in NBCCs, BCC | Hahn et al (1996). *Cell* | Yes | 1 | 1 | 2 | 1 |
| | Johnson et al (1996). *Science* | Yes | 2 | 2 | 4 | 2 |
| 1996: Ptch is a Hh receptor | Stone et al (1996). *Nat* | Yes | 4 | 4 | 6 | 5 |
| 1996: Cyclopia by impaired Hh signaling | Chiang et al (1996). *Nat* | Yes | 6 | 5 | 1 | 6 |
| 1996: HH pathway mutations in human cyclopia | Belloni et al (1996). *Nat Genet* | Yes | | | 7 | 9 |
| | Roessler et al (1996). *Nat Genet* | Yes | 10 | | 3 | 7 |
| 1997: First transgenic BCC mouse model | Oro et al (1997). *Science* | No | | | | |
| 1997: *Ptch1* mutant mice phenocopy NBCCs | Goodrich et al (1997). *Science* | No | | | | |
| 1998: Cyclopamine inhibits Hh signaling | Cooper et al (1998). *Science* | Yes | | 2 | | 2 |
| 1998: SMO mutations in BCC | Xie et al (1998). *Nature* | Yes | | 1 | | 1 |
| 2000: Cyclopamine inhibits oncogenic Hh signaling | Taipale et al (2000). *Nature* | Yes | | | 1 | |
| 2002: First synthetic small molecule Hh pathway inhibitor | Chen et al (2002). *Genes Dev* | Yes | 1 | | 1 | 1 |
| 2003: Multiple cancers with aberrant Hh signaling | Grachtchouk et al (2003). *EMBO J* | No | | | | |
| 2003: SMO antagonist pre-clinical BCC Tx | No Document Referenced | No | | | | |
| 2004: SMO antagonist pre-clinical medulloblastoma Tx | No Document Referenced[d] | Yes | | 2 | 7 | |
| 2009: Phase I trials of systemic Hh pathway inhibitor | Rudin et al (2009). *N Engl J Med* | Yes | 2 | 2 | 2 | 2 |
| | Von Hoff et al (2009). *N Engl J Med* | Yes | 1 | 1 | 1 | 3 |
| | Yauch et al (2009). *Science* | Yes | 3 | 3 | 3 | 1 |
| 2011: Topical Hh pathway inhibitor BCC trials | Skvara et al (2011). *J Invest Dermatol* | No | | | | |
| | Tang et al (2011). *Clin Cancer Res* | No | | | | |
| 2012: Hh inhibitor vismodegib FDA approved | No Document Referenced[e] | Yes | | 1 | | |
| 2012: Systemic Hh inhibitor BCC prevention trial | Tang et al (2012). *N Engl J Med* | Yes | | 2 | | |

[a] Wong and Dlugosz do not include a reference to any articles published in the 1950s. They do, however, reference a 1978 review by Keefer, who in turn cites Binns et al (1959) to describe operators experiencing up to 25% deformed lambs in bands of ewes from areas of Idaho.
[b] Though Wong and Dlugosz list 1968 as a milestone which is detected by RPYS, they do not reference the 1968 article on cyclopamine by Keeler and Binns. They do, however, cite Keeler and Binns (1966), which identified three alkaloids potentially inducing cyclopia.
[c] Wong and Dlugosz include 1993 as a milestone which is detected by RPYS. They do not reference any 1993 article in their bibliography, however, RPYS analysis identifies an article cloning of a family of mammalian hedgehog genes by Echelard et al (1993) as the most referenced paper in 1993.
[d] Wong and Dlugosz include 2004 as a milestone which is detected by RPYS. However, they do not reference any 2004 article in their paper. RPYS analysis of the second WoS search identifies an article by Romer et al (2004) on suppressing sonic hedgehog pathway to eliminate medullablastomas in mice.
[e] FDA approval follows positive results from clinical trials. While there was no formal citation for FDA approval, Wong and Dlugosz do cite the 2012 publication of Clinical Trials on Vismodegib by Sekulic et al. (2012).

**Table 2.** Performance of RPYS in detecting seminal articles compared to key research breakthroughs described by subject matter experts. The leftmost column, Milestones, provides the year and description of key milestones noted in Wong and Dlugosz (2014; figure 2). The next column, Milestone Articles, maps

articles from the bibliography or text of Wong and Dlugosz to the milestones they describe. We note that for some occasions multiple papers map to milestones. The column RPYS Peak reveals whether or not RPYS results of searches A, B, C or D yield a *peak* in the spectrogram corresponding to that milestone year. Finally, the last set of columns, Rank in Search, shows for each search A-D whether the milestone articles underlying an RPYS peak was in the top 1-10 most cited references for that year.

Further, RPYS aided in the clarification of seminal research underlying some of the milestones discussed in Wong and Dlugosz that were not explicitly cited (see Table 2 footnotes for more details). For example, Wong and Dlugosz note the 1950s occurrence of "one eyed lambs in Idaho" as a seminal event for research into the underlying mechanisms of basal cell carcinoma. However, the article does not cite any specific works published in the 1950s. It does, however, refer to a review article on this topic:

> The identification of the first pharmacological inhibitor of Hh signaling was based on a series of pivotal observations and discoveries dating back to the 1950s (Figure 2), when up to 25% of lambs on certain Idaho ranches were born with severe craniofacial deformities, including a single eye (cyclopia) (reviewed in Keeler, 1978).

Examining the first paragraphs of the cited review by Keeler (1978) reveals the following description:

> During the first half of the century, a significant percentage of lambs born in certain parts of Idaho were congenitally deformed (1-3)… By the 1950s some operators were experiencing an incidence of up to 25% deformed lambs in bands of ewes from these areas (1-3).

In this section, Keeler cites a 1959 article entitled "A congenital cyclopean-type malformation in lambs" by Wayne Binns et al. This same article is the most referenced article underlying a peak in 1959 in the spectrogram for Search C. In such cases, we consider the results of RPYS to have successfully captured a milestone article.

**Conclusions**

We demonstrate the application of new analytic techniques to connect major scientific discoveries with medical innovation. We show that multi-RPYS and standard RPYS, a relatively new class of citation algorithms, successfully capture key milestones in a field. We demonstrated that the combination of multi-RPYS and inferential statistics far exceeded chance expectations for identifying milestone years (70% success rate compared with less than 10% expected by chance). Then, we showed that 76% of the milestone papers underlying milestone years discussed by experts from the field of cutaneous malignancy are easily identifiable using online-available standard RPYS tools. This offers the possibility that recently developed RPYS tools (Comins and Leydesdorff 2016b; Leydesdorff et al. 2014; Thor et al. 2016) could empower a broader array of researchers to discover seminal works in scientific areas.

We wish to emphasize that expert opinion is necessary for contextualizing and describing the seminal works; but RPYS can expedite the process of *finding* milestones in the literature. In short, we conclude that RPYS represents a significant advance for maturing citation algorithms to identify seminal works in scientific fields, a key goal in scientometrics.

The performance achieved by these algorithms also opens the door for broader use of data-driven tools for mapping major milestones in the history of medical innovation (such as the prototype described in Comins & Leydesdorff, 2016b). Indeed, with the astounding pace of scientific growth (Bornmann & Mutz, 2015) and the tendency for fields with rapidly changing research fronts to cite more recent literature (Leydesdorff et al., 2014), such tools provide an easier way to identify key scientific activities influencing areas of medical innovation. RPYS and multi-RPYS offer a general purpose method to revive the appreciation of early scientific achievements for the next generation of scientists and engineers, policy makers, funding agency program managers as well as lay people alike.